\documentclass[manuscript]{aastex}
\usepackage{emulateapj5}
\usepackage{graphicx}
\usepackage{times}

\renewcommand{\topmargin}{0.1cm}

\shortauthors{A. M. Hopkins}
\shorttitle{On the evolution of star forming galaxies}

\begin{document}

\title{On the evolution of star forming galaxies}

\author{A. M. Hopkins\altaffilmark{1,2}
}

\affil{
\begin{enumerate}
\item Dept.\ of Physics and Astronomy, University
 of Pittsburgh, 3941 O'Hara Street, Pittsburgh, PA 15260
\item Hubble Fellow; email ahopkins@phyast.pitt.edu
\end{enumerate}
}

\begin{abstract}
The evolution in the comoving space density of the global average
galaxy star formation rate (SFR) out to a redshift around unity is well
established. Beyond $z\approx1$ there is growing evidence that this
evolution is flat or even increasing, contrary to early indications
of a turnover. Some recent analyses of $z\approx6$ photometric
dropouts are suggestive of a decline from $z=3$ to $z\approx6$, but
there is still very little constraint on the extent of dust obscuration at
such high redshifts. In less than a decade, numerous measurements of
galaxy SFR density spanning $z=0$ to as high as $z\approx6$ have rapidly
broadened our understanding of galaxy evolution, and a summary of existing
SFR density measurements is presented here. This global star formation
history compilation is found to be consistent to within factors of about
three over essentially the entire range $0<z\lesssim6$, and it can be used
to constrain the evolution of the luminosity function (LF) for star
forming (SF) galaxies. The LF evolution for SF galaxies has been previously
explored using optical source counts, as well as radio source counts at
1.4\,GHz, and a well-known degeneracy between luminosity evolution
($L\propto(1+z)^Q$) and density evolution ($\phi\propto(1+z)^P$) is found.
Combining the constraints from the global SFR density evolution with those
from the 1.4\,GHz radio source counts at sub-millijansky levels allows this
degeneracy to be broken, and a best fitting evolutionary form to be
established. The preferred evolution in a
$H_0=70,\Omega_M=0.3,\Omega_\Lambda=0.7$ cosmology from these combined
constraints is $Q=2.70\pm0.60$, $P=0.15\pm0.60$.
\end{abstract}

\keywords{galaxies: evolution --- galaxies: starburst ---
 radio continuum: galaxies}

\section{Introduction}
\label{int}

The increase by an order of magnitude in the global comoving space density
of galaxy star formation rate (SFR) from $z=0$ to $z\approx1$ has been well
established by numerous measurements in recent years
\citep[e.g.,][see also the list of references in Tables~\ref{datatab} and
\ref{lanztab}]{Wil:02,Haa:00,Flo:99,Cow:99,Hog:98,Ham:97,Lil:96}.
A large body of data has now been collected to measure this evolution,
and the shape of the star formation (SF) history at higher redshifts
(now approaching $z\approx6$). As will be seen, within measurement
uncertainties and the limitations of the various surveys used, this large,
heterogeneous dataset is highly consistent over the entire redshift
range $0<z\lesssim6$, and constrains the SFR density to within a factor of
about three at most redshifts.

With such a wealth of measurements now available, the evolution of
the global galaxy SFR density can be used as a robust constraint on
various simulations and semianalytic models of galaxy evolution. This
is already being pursued by many authors \citep[e.g.,][]{Som:01,Pei:99},
and tantalising indications about the details of galaxy evolution are
suggested as a result \citep[such as collisionally induced star formation
being favoured as the dominant mode of galaxy evolution,
as opposed to steady ongoing star formation,][]{Som:01}. The richness of
the available measurements offers a very strong constraint that must
be met by such models, but full advantage has not yet been taken of
this important resource.

To illustrate this point, a compilation drawn from the literature of SFR
density measurements as a function of redshift is presented here, and
constraints on the evolving luminosity function of star forming galaxies
are derived. Together with constraints from radio source
counts, the SF history data allow the degeneracy between
rates of luminosity and density evolution to be broken, and a robust
constraint on both is derived.

The details of the measurements from the literature
are presented in \S\,\ref{meas}, and evolving luminosity functions
for star forming galaxies are presented in the context of the SF history
diagram in \S\,\ref{evollf}. In \S\,\ref{lfconstraint} the SF
history compilation is used in conjunction with the radio source
counts to constrain the form for the evolving LF. The results are
discussed in \S\,\ref{disc}, and conclusions summarised in \S\,\ref{summ}.
A flat Lambda cosmology is assumed throughout, as indicated by
numerous recent measurements, with $H_0=70\,$km\,s$^{-1}$\,Mpc$^{-1}$,
$\Omega_M=0.3$, $\Omega_{\Lambda}=0.7$ \citep[e.g.,][]{Spe:03}.

\section{The measurements}
\label{meas}

Measurements of galaxy SFR density have been compiled from the literature,
converted to a common cosmology and SFR calibration, and corrected
for dust obscuration, where necessary, in a common fashion.
This is necessary in order to consistently compare measurements made at
different wavelengths. The star formation calibrations chosen are presented in
Table~\ref{sfrcal}, all assuming a \citet{Sal:55} initial mass function (IMF)
and mass range $0.1<M<100\,M_{\odot}$. The H$\alpha$, UV and far infrared (FIR)
calibrations adopted are those from \citet{Ken:98}. The 1.4\,GHz SFR
calibration chosen is that of \citet{Bell:03}, derived to be consistent
with the FIR calibration from \citet{Ken:98}. This calibration gives
SFRs lower by about a factor of two than the calibration of \citet{Con:92}
for $\log(L_{1.4}/{\rm W\,Hz}^{-1})\gtrsim21.8$. As luminosity decreases the
SFRs approach those of \citet{Con:92}, and become higher for luminosities
below $\log(L_{1.4}/{\rm W\,Hz}^{-1})\lesssim20.5$.

\subsection{Cosmology conversion}
\label{cosmoconv}

The conversion from the originally chosen cosmology to that assumed
here is as follows. Since in a flat universe comoving
volume is proportional to comoving distance cubed, $V_c \propto D_c^3$,
the comoving volume between redshifts $z-\Delta z$ and $z+\Delta z$,
$V_c(z,\Delta z) \propto D_c^3(z+\Delta z) - D_c^3(z-\Delta z)$.
Since luminosity is proportional to comoving distance squared,
$L \propto D_c^2$, then the SFR density for a given redshift range,
(assuming that all galaxies lie at the central redshift),
\begin{equation}
\dot{\rho}_* \propto \frac{L(z)}{V_c(z,\Delta z)} \propto
 \frac{D_c^2(z)}{D_c^3(z+\Delta z) - D_c^3(z-\Delta z)}.
\end{equation}
So, in order to convert from one cosmology to another, this expression
is evaluated for both cosmologies at the appropriate redshift and the
ratio used as the conversion factor.
This is similarly described by \citet{Asc:02}, and an equivalent
form used by \citet{Hog:02}.

In the case of converting luminosity functions (LFs), the conversion of the
luminosity and volume components need to be done separately.
When applying a luminosity-dependent obscuration correction, as
described below, the LF being used must first be
corrected for the cosmology to ensure that the obscuration corrections
applied correspond to the correct luminosities in the final cosmology.
To convert the LF, the characteristic luminosity and the
volume normalisation factors ($L^*$ and $\phi^*$ in the \citet{Sch:76} or
\citet{Sau:90} parameterisations) are converted. This is done in a
similar fashion to that described above, using the proportionalities
already given.

This procedure for converting cosmologies, with the coarse assumption
that all galaxies lie at the central redshift, is only approximate.
The expected uncertainties introduced by this assumption, though, even in
the case of the measurements spanning the largest redshift ranges, are
at most of the order of $10\%$, which is small compared to the factors of two
to three by which the actual measurements may differ from each other,
as seen below.

\subsection{Obscuration corrections}
\label{obscor}

Obscuration by dust is well known to affect measurements of galaxy
luminosity at UV and optical wavelengths. Correcting for this effect
is not always straightforward, however, since measurements of an
obscuration sensitive parameter, such as the Balmer decrement or
UV spectral slope, are not always easy to obtain. It has become
a standard procedure to make an approximate obscuration correction
by assuming an average level of expected obscuration
(such as $A_V=1\,$mag) and uniformly correcting all objects. This procedure
has the advantage that it is straightforward to apply either to a luminosity
function or a luminosity density, being simply a scaling factor.
Recently, though, it has been shown that galaxies with high luminosities
or SFRs tend, on average, to suffer greater obscuration than faint
or low SFR systems
(\citeauthor{Hop:01} \citeyear{Hop:01}; \citeauthor{Sul:01} \citeyear{Sul:01};
\citeauthor{Per:03} \citeyear{Per:03}; \citeauthor{Afo:03} \citeyear{Afo:03};
\citeauthor{Hop:03} \citeyear{Hop:03}; but see also
\citeauthor{Buat:02} \citeyear{Buat:02}).
Correcting for such a luminosity-dependent (or SFR-dependent) effect is
not as straightforward as for a common obscuration, though, since
a knowledge of the full LF is necessary. It is further
complicated by the fact that the specific relation describing the
luminosity-dependent effect is strongly dependent on the selection
criteria of the sample being investigated \citep{Afo:03,Hop:03}.

These issues are not insurmountable, however, and both a common obscuration
and a luminosity-dependent obscuration are explored below. In some cases
actual measurements of the obscuration (such as the Balmer decrement) have
already been used by the original authors, with obscuration curves consistent
with those assumed below, to correct the LFs or the SFR density estimate.
These results are always used in favour of invoking less reliable average
corrections. It must also be emphasised that although different average
obscurations are assumed below for the common obscuration correction,
depending on the wavelength used to select the sample being corrected,
these are not necessarily inconsistent. They are based on observed average
obscurations for similarly selected samples in the literature. Some issues
regarding the use of average obscuration corrections are discussed further
below, but it should be clear that for surveys that detect only relatively
low obscuration systems, for example, a relatively small average obscuration
correction is appropriate. The corresponding LF derived from such a
sample is then used (by fitting a Schechter function, for example) to
infer the details of brighter but more heavily obscured systems that fall
below the detection threshold, as well as systems below the survey limit
that are fainter and possibly less obscured.

In all obscuration corrections used herein for emission line SFR
measurements (H$\alpha$, H$\beta$, [O{\sc ii}]) the galactic
obscuration curve from \citet{Car:89} is used. For continuum
measurements (primarily UV wavelengths from 1500\,\AA\ to 2800\,\AA)
the starburst obscuration curve from \citet{Cal:00} is used \citep[see
also][]{Cal:01}. When applying a common obscuration correction, $A_{\rm
H\alpha}=1.0\,$mag is assumed for emission line measurements, often
found as the average obscuration to H$\alpha$ emission in samples
of local galaxies \citep[e.g.,][]{Ken:92,Hop:03}. Similarly, for
UV-selected samples a typical obscuration is found to be $A_V=1.0\,$mag
\citep[e.g.,][]{Sul:00,Tres:96}, although this value is based on galactic
obscuration curves rather than the starburst curves from Calzetti assumed
here. To convert this to the appropriate $A_V$ for the Calzetti curve, the
mean Balmer decrement corresponding to this value is first inferred. The
analysis of \citet{Sul:00} is used as a reference here, and those
authors base their obscuration corrections on the \citet{Sea:79} galactic
obscuration curve. Since $R = A_V/E(B-V)$ and $R=3.2$ is assumed, the
mean value of $A_V=1.0$ \citep[from][]{Sul:00}, gives $E(B-V)=0.313$. But
$E(B-V)$ also depends on the choice of obscuration curve,
\begin{equation}
\label{ebv}
E(B-V) = \frac{\log[(f({\rm H\alpha})/f({\rm H\beta}))/2.86]}
              {0.4(k({\rm H\beta}) - k({\rm H\alpha}))}.
\end{equation}
\citep{Cal:96}. Here $f({\rm H\alpha})/f({\rm H\beta})$ is the Balmer
decrement, and $k({\rm H\beta}) - k({\rm H\alpha})$ is the difference between
the values of the obscuration curve at the wavelengths of H$\beta$ and
H$\alpha$. The latter quantity is 1.19 for the \citet{Sea:79} curve, and
gives a Balmer decrement of $4.03$ corresponding to the mean $A_V=1$.
For the \citet{Cal:00} curve, $k({\rm H\beta}) - k({\rm H\alpha})=1.27$, and
this Balmer decrement gives $E(B-V)_{\rm gas}=0.293$, similar to the above
value. Indeed, $A_{V\rm gas}=k(V) E(B-V)_{\rm gas}=1.19$ (with $k(V)=4.05$), a
little higher than the 1\,mag from the galactic curve. To correct the stellar
continuum the relation $E(B-V)_{\rm star}=0.44 E(B-V)_{\rm gas}$
is used \citep{Cal:00}. The subscript notation ``gas" and ``star" refer
to nebular and stellar continuum obscuration respectively. Finally,
$A_{V\rm star}=0.44 k(V) E(B-V)_{\rm gas} = 0.52$. Note that
in Equation~\ref{ebv} it is necessary to use $k(\lambda)$ appropriate
for nebular obscuration \citep[Equation~4 of][]{Cal:00}.
Hence, $A_V=1.0$ for the \citet{Sea:79} curve becomes $A_{V\rm star}=0.52$
for the \citet{Cal:01} curve. It would simplify comparisons such as this
to have obscuration sensitive quantities presented as observables,
such as Balmer decrement or UV spectral slope. These should be elementary to
provide in addition to derived values dependent on the chosen obscuration
curve, like $A_V$ and $E(B-V)$.

To summarise, when a common obscuration correction is performed below,
emission line measurements are corrected using $A_{\rm H\alpha}=1.0$
and the \citet{Car:89} galactic obscuration curve, and continuum
UV measurements are corrected using $A_{V\rm star}=0.52$ and the
\citet{Cal:00} starburst obscuration curve.

It is interesting to note that an obscuration at the wavelength of
H$\alpha$ of $A_{H\alpha}=1.0\,$mag corresponds to $A_V=1.22$, using the
galactic obscuration curve of \citet{Car:89}, or $A_V=1.29$ using the
curve of \citet{Sea:79}, compared to $A_V\approx1.0$ (using galactic
obscuration curves) found for UV selected samples. This observation
is consistent with the expected result that the average obscuration
of galaxies selected at redder wavelengths (around H$\alpha$) can be
greater than those selected at UV wavelengths where the more heavily
obscured systems will not enter the sample. This effect is seen quite
strikingly in the radio-selected sample of star forming galaxies analysed
by \citet{Afo:03} who find systematically higher median obscurations
at all luminosities than in the optically selected sample explored by
\citet{Hop:03}.

The analysis of \citet{Mas:01}, in addition, emphasises the important
result that making obscuration corrections based on an average observed
$A_V$ or $E(B-V)$ will result in an {\em underestimate\/} of the true
measurement. They explain that the {\em effective\/} mean obscuration
is larger than the mean of the observed obscurations. This implies that
{\em the SFR densities obtained using the common obscuration corrections
above will be underestimates of the true SFR densities}.

The analyses of SFR-dependent obscuration \citep{Hop:01,Sul:01,Afo:03,Hop:03}
clearly indicate that using such empirical trends to make obscuration
corrections should only be done in the absence of more direct measurements.
In the case of measurements in the literature that have been obscuration
corrected directly using Balmer decrements and obscuration curves consistent
with those adopted here, these obscuration corrected results are
taken directly (with appropriate cosmology and SFR calibration conversions
where necessary). Where no obscuration correction has been made but an
observed LF published, an SFR-dependent obscuration
correction can be performed. For the form of this correction herein,
the relation between obscuration and SFR from \citet{Hop:01} is adopted
(after conversion to the presently assumed cosmology). This relation is
illustrated briefly for obscuration at the wavelength of H$\alpha$, as
follows. Objects with $\log(L_{H\alpha}/{\rm W})<33$, or equivalently
${\rm SFR}_{H\alpha}<0.08\,M_{\odot}$\,yr$^{-1}$, are assumed to suffer no
obscuration. Above this level the effect of obscuration increases to, e.g.,
a factor of 2 at an observed (prior to obscuration correction)
$\log(L_{H\alpha}/{\rm W})=34.1$ corresponding to an apparent
${\rm SFR}_{H\alpha}=1\,M_{\odot}$\,yr$^{-1}$. This increases again to a
factor of 5 at an observed $\log(L_{H\alpha}/{\rm W})=36.4$, or
${\rm SFR}_{H\alpha}=198\,M_{\odot}$\,yr$^{-1}$, and continues increasing
indefinitely for higher luminosities
\citep[compare with Figures~2a and 3a of][]{Hop:01}.

It is important to note that this form appears to be appropriate for
optically selected samples, particularly those restricted to higher
equivalent width systems. A similar relation was found by \citet{Sul:01},
although \citet{Per:03} finds a relation with a steeper slope, and
\citet{Hop:03} show that when lower equivalent width systems enter the
sample the median obscurations for a given luminosity or SFR can be
notably higher. In a radio-selected sample, free from obscuration based
selection biases, \citet{Afo:03} find a relation with significantly higher
median obscurations for a given SFR compared to optically selected samples
\citep[see discussion in][]{Hop:03}. Since the samples to be corrected
herein are UV or H$\alpha$ selected, the form from \citet{Hop:01} was
deemed appropriate. It should be noted that if other published relations
were used, this would result in {\em larger\/} obscuration corrections
than applied here. If a galactic obscuration curve is used for the UV
continuum corrections rather than that of \citet{Cal:00}, the obscuration
corrections for the UV SFR density measurements would again be larger.

\subsection{[O{\sc ii}] SFR calibrations}
The issue of SFR calibrations for [O{\sc ii}] luminosities is a complex
one. In particular, the [O{\sc ii}] to H$\alpha$ flux ratio of
0.45 \citep{Ken:92,Ken:98} for local galaxies is now recognised
to be strongly luminosity-dependent \citep{Jan:01}, as well as having
metallicity and obscuration dependencies. \citet{Ara:02} compared
observed and obscuration corrected [O{\sc ii}]/H$\alpha$ line flux ratios
in two different local galaxy samples, the H$\alpha$ selected
Universidad Complutense de Madrid (UCM) survey \citep{Gal:96} and
the $B$-band selected Nearby Field Galaxy Survey, NFGS, \citep{Jan:01}.
They find that the luminosity dependence is primarily due to obscuration
effects, and for the UCM survey the obscuration corrected
[O{\sc ii}]/H$\alpha$ ratio has considerably smaller scatter than the
observed ratio, with a median value close to unity. The obscuration corrected
ratio is also largely independent of luminosity, although they caution that
sample selection effects still need to be carefully accounted for.
\citet{Tres:02} explore the [O{\sc ii}]/H$\alpha$ flux ratio for
different redshift ranges using the Canada-France Redshift Survey (CFRS).
They find little evidence for evolution in the luminosity dependence
of the observed flux ratio out to $z\approx 1$. The important conclusion
is that an [O{\sc ii}]/H$\alpha$ flux ratio appropriate to the sample,
given the sample selection effects, should be chosen when converting
[O{\sc ii}] luminosities to H$\alpha$ and to SFR.

Of the [O{\sc ii}]-based SFR density estimators compiled here,
\citet{Tep:03} adopts the relation of \citep{Jan:01} in deriving
the SFR density at $z=0.9$, and they comment that the average value
of the [O{\sc ii}]/H$\alpha$ ratio is close to 0.45 for their sample.
\citet{Ham:97} presents observed [O{\sc ii}] luminosity densities
for the CFRS sample. Over the redshift range spanned by these measurements,
\citet{Tres:02} find a median observed [O{\sc ii}]/H$\alpha$ ratio of
0.46 for CFRS galaxies. This ratio is adopted here in converting the
[O{\sc ii}] luminosity densities of \citet{Ham:97} to SFR densities.
The [O{\sc ii}]-derived SFR densities from \citet{Hog:98} assume
the local ratio of 0.45, although this is not constrained by independent
estimates for the sample used. Since higher luminosities will dominate
the sample at higher redshifts, lower [O{\sc ii}]/H$\alpha$
flux ratios may be more appropriate at high redshifts. This would act
to steepen the slope of the SFR density with redshift defined by these points.

\subsection{The compilation}
\label{compilation}
\begin{figure*}
\centerline{\rotatebox{-90}{\includegraphics[width=10cm]{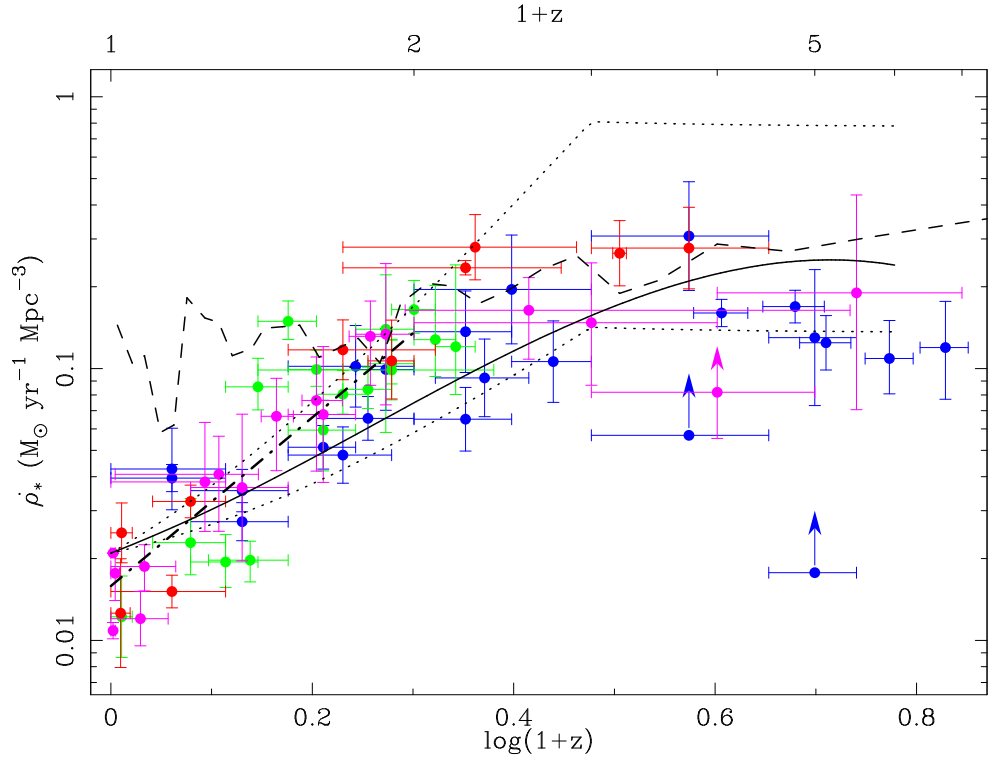}}}
\caption{Evolution of SFR density with redshift, using a common obscuration
correction where necessary. The points are color-coded by rest-frame
wavelength as follows. Blue: UV; Green: [O{\sc ii}]; Red: H$\alpha$ and
H$\beta$; Pink: X-ray, FIR, sub-mm and radio.
The solid line is the evolving 1.4\,GHz LF derived by \citet{Haa:00}.
The dot-dashed line shows the least squares fit to all the $z<1$ data
points, $\log(\dot{\rho}_*) = 3.10\log(1+z) - 1.80$. The dotted lines show
pure luminosity evolution for the \citet{Con:89} 1.4\,GHz LF, at
rates of $Q=2.5$ (lower) and $Q=4.1$ (upper). The dashed line is
the ``fossil" record from Local Group galaxies \citep{Hop:01b}.
 \label{fig1}}
\end{figure*}

\begin{figure*}
\centerline{\rotatebox{-90}{\includegraphics[width=10cm]{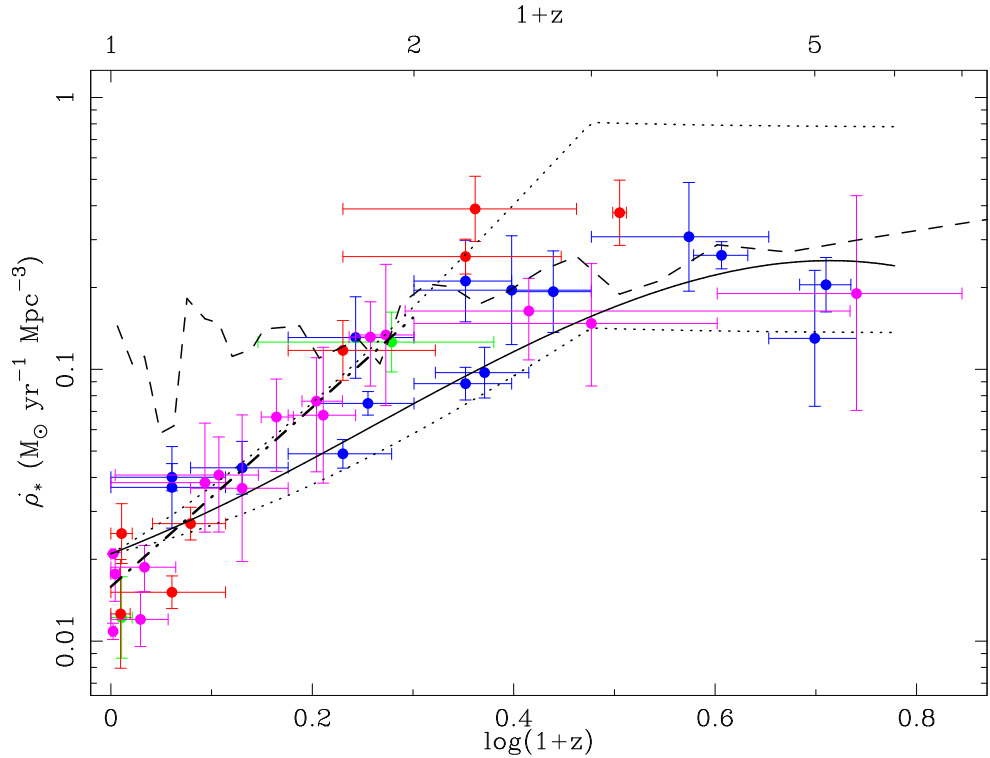}}}
\caption{As for Figure~\ref{fig1}, with a luminosity-dependent obscuration
correction. Symbols and lines as in previous Figure. The least squares
fit to the $z<1$ data points (dot-dashed line) in this Figure is
$\log(\dot{\rho}_*) = 3.29\log(1+z) - 1.80$.
 \label{fig2}}
\end{figure*}

The collection of measurements from the literature is presented in
Table~\ref{datatab}, with $\dot{\rho}_*$ given for both methods of
obscuration correction, common and luminosity-dependent. The same
values appear in both these columns for measurements not requiring
obscuration correction (IR, sub-mm and radio), or for those where
Balmer decrement measurements were used directly by the original
authors to correct the data prior to deriving luminosity functions.
This table also indicates the effective factors used in performing the
cosmology correction from the original reference, and in both types of
obscuration correction. A total of 33 references provide data here, giving
a total of 66 points measured in the $(z, \dot{\rho}_*)$ plane. These are
shown in Figure~\ref{fig1}. Of these points, 41 have measured obscuration
corrections, are able to have an SFR-dependent obscuration correction
applied, or have no need of obscuration corrections, and these results
are shown in Figure~\ref{fig2}. Adjacent data points from the [O{\sc ii}]
derived measurements of \citet{Hog:98} are not independent, having been
constructed in overlapping bins.

The LF parameters, where available, are given in Table~\ref{lfparams} after
converting to the cosmology assumed here. This table indicates whether the
published LF includes obscuration corrections based on observed obscuration
estimates or not. If not, integrations over these LFs are performed in order
to apply the SFR-dependent obscuration correction detailed above. A fixed
luminosity range is used for these integrations, depending on the wavelength
at which the LF is estimated. For H$\alpha$ LFs, the integration is over
$30.0\le\log(L_{\rm H\alpha}/{\rm W})\le40.0$. For UV LFs the range is
$15.0\le\log(L_{\rm UV}/{\rm W\,Hz^{-1}})\le25$. For 1.4\,GHz LFs the
range is over $15.0\le\log(L_{\rm 1.4}/{\rm W\,Hz^{-1}})\le28.0$.
The data from the Great Observatories Origins Deep Survey
\citep[GOODS,][]{Gia:04}, which include the two highest redshift UV-based
data points, come from an integration of their derived luminosity functions
that has an imposed absolute magnitude cutoff brighter than the limits assumed
here for the other surveys. As a result the points corresponding to this
survey shown in Figure~\ref{fig1} are actually lower limits to the total SFR
density that would have been derived with the luminosity range indicated above.
Integration of the LF over a magnitude range comparable with the above
luminosity range is likely to increase these estimates by a factor of
two or so. This would improve the level of consistency with the high redshift
sub-mm measurement, and the evolving radio luminosity function results
(see \S\,\ref{evollf} below). It is worth emphasising here, as an aside,
that many high-redshift studies of the comoving SFR density (some of which are
discussed further in \S\,\ref{disc2} below) impose a luminosity or magnitude
limit on the integration of the LF, to avoid strong biases from the assumption
of a faint end slope poorly constrained by observation. When comparing
measurements in the literature it is obviously desirable to be consistent
in the extent of the LF integrations being compared as, depending on the
limits chosen, the results can vary by factors of two to four or so.

A common parameterisation for the evolution of the SFR density with
redshift is a simple power law, $\dot{\rho}_* \propto (1+z)^\beta$, for
$z\lesssim 1$, where $\beta$ has been estimated in the literature to range
from around 1.5 \citep{Cow:99} to 3.9 \citep{Lil:96}, with most favoured
values around 3, for an Einstein-de Sitter (EdS) cosmology.
The ``meta-analysis" of \citet{Hog:02} gives a robust measurement of
of $\beta=3.3\pm0.8$ for an EdS cosmology, and a weighted
mean fit of $\beta=2.74\pm0.28$ for the cosmology assumed here.
Using the current data compilation, restricted to measurements for $z\leq1.0$,
an ordinary least squares \citep[OLS,][]{Iso:90} regression
of $\log(\dot{\rho}_*)$ on $\log(1+z)$ is performed to estimate both
$\beta$ and $\log[\dot{\rho}_*(z=0)]$. For the assumption of a common
obscuration this gives
\begin{equation}
\log(\dot{\rho}_*) = (3.10\pm0.25) \log(1+z) - (1.80\pm0.04).
\end{equation}
\citep[Although both overlapping binnings for the data of][are shown in
Figure~\ref{fig1}, only the independent bins at $z=0.2, 0.4, 0.6, 0.8, 1.0$
have been used in this fit.]{Hog:98}
This result is consistent with that of \citet{Hog:02}, although
somewhat higher. The main reason for this is that the measurements of
\citet{Mob:99} have been neglected in the current compilation. If that
result is omitted from the analysis of \citet{Hog:02}, a value of
$\beta=3.15\pm0.29$ would have been derived, highly consistent
with the present estimate. The measurements of \citet{Mob:99}
are omitted here primarily because of the significant incompleteness
affecting their estimates, particularly for the higher redshift bin.
For the case of the luminosity-dependent obscuration, the OLS
regression gives
\begin{equation}
\log(\dot{\rho}_*) = 3.29\pm0.26\log(1+z) -1.80\pm0.04.
\end{equation}
The consistency of these two results should not be too surprising,
given that the 1.4\,GHz and FIR derived data remain unchanged from
one to the other. Both these relations are presented as dot-dashed lines in
Figures~\ref{fig1} and \ref{fig2} respectively. It is worth emphasising the
similarity, though, between many of the data points at $z<1$ when corrected
for obscuration using the two different methods. For these points the
difference between the two different methods ranges from $\approx -20\%$
\citep{Trey:98,Tres:98}, where the SFR dependent obscuration corrected value
is lower than with the common obscuration, to $\approx +30\%$
\citep[the point at $z=0.75$]{Con:97}.

\subsection{Surface brightness dimming}

The question of whether any given survey is complete over the full
range of galaxy sizes for each interval of fixed luminosity
is complex to address comprehensively. In most cases flux (or flux
density) limits imposed in constructing source catalogues are effectively
surface brightness thresholds \citep[witness the so-called ``resolution
correction" made to radio galaxy source counts, e.g.,][when constructed
from peak flux density selected samples, to account for missing low
surface brightness sources]{Hop:03a}. Different methods used by different
authors to address (or ignore) this issue, which affects both low and high
redshift samples, may contribute to some extent to the scatter in the
values seen for the estimated SFR density at each redshift.

\begin{figure*}
\centerline{\rotatebox{-90}{\includegraphics[width=10cm]{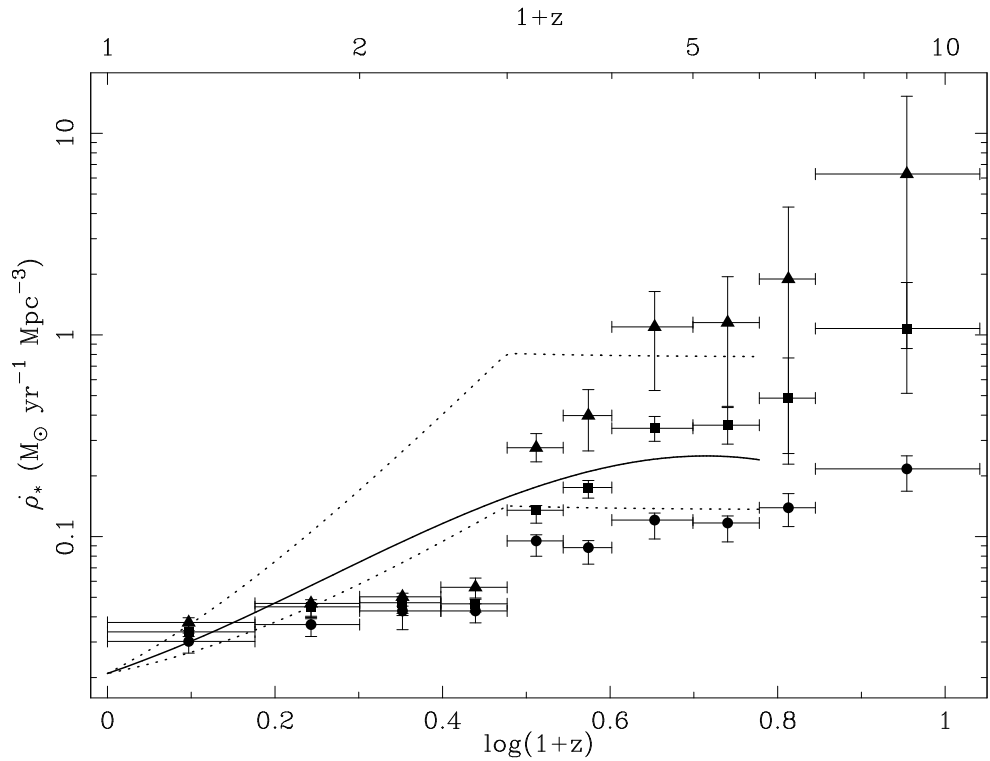}}}
\caption{Lines are as in previous Figures, although note the different
scales on the axes. The points show the data from Figure~4 of
\citet{Lan:02} (circles correspond to the green data points of that Figure,
squares to the blue points and triangles to the red points), after correction
for a common obscuration. This emphasises the high level of consistency
between the current compilation and the UV estimates at high redshift, even
when the effects of surface brightness dimming are taken into account.
 \label{fig3}}
\end{figure*}

Photometric redshifts have been used with large, deep samples of galaxies to
estimate the SF history consistently over a very broad redshift range
\citep{Pas:98,Lan:02}. The issue of correctly accounting for surface
brightness dimming is of particular concern with this method, as a fixed
flux limit corresponds to dramatically different surface brightness limits at
different redshifts, and this can strongly bias the inferred SF history
results. This problem is explored in detail by \citet{Lan:02}, in which the
SFR intensity distribution, $h(x)$, is introduced. The evolution of $h(x)$
is explored by assuming three different models, one corresponding to
luminosity evolution, one to a surface-density evolution, and
an intermediate model defined by allowing the break point of the
double-power-law distribution to evolve. The results of that analysis
are shown in Figure~\ref{fig3}, and presented in Table~\ref{lanztab},
after incorporating obscuration corrections using the common obscuration
assumptions described above. It can be seen that the three models explored
by \citet{Lan:02} are highly consistent with the range of values measured
in the other studies compiled here. The central model seems to be most
consistent, and this is also apparently the most consistent with the
data derived in \citet{Lan:02} from the neutral hydrogen column density
measured from damped Ly$\alpha$ absorption systems between redshifts
$2.5<z<5$. There is also an apparent underestimate for all three models
around $z\approx1.5$ compared to the numerous other measurements in
this region, as well as when compared to the lower limit inferred
from the radio source counts, (lower dotted line, see \S\,\ref{evollf}
below). There is also a continual increase above $z\approx5$, counter
to the assumptions made in the evolving radio LF modelling below. It is
possible that the photometric redshift estimation erroneously identifies
extreme redshifts (higher than $z\approx6$ or 7) for objects whose
true redshifts are around $z\approx1.5$, where the 4000\,\AA\ break
(an important feature in constraining photometric redshifts) is shifted
out of the optical spectral window.

Apart from this issue, the SFR density evolution measured from photometric
redshift estimates, when corrected for obscuration effects, appears highly
consistent with all the other measurements and with the constraints from
radio source counts. This is an encouraging result, suggesting that possible
surface brightness related biases in the other estimates of SFR density
compiled here are at most similar to the scatter seen between different
measurements at similar redshifts. They may well be somewhat less given
the other sources of uncertainty involved. This does not suggest that
these effects should be ignored, but rather that more or less appropriate
corrections have been used by most authors in accounting for these effects.

\section{Evolving luminosity functions}
\label{evollf}
In order to have a reference for interpreting these results and the
overall compilation, evolving LFs for star forming (SF) galaxies have
been explored.

Numerous measurements of evolutionary parameters or forms for galaxy
LF exist in the literature. For the current exploration the evolving
1.4\,GHz LF for star forming (SF) galaxies from \citet{Haa:00} is
particularly relevant, as are the constraints on 1.4\,GHz SF LF evolution
from deep radio source counts \citep[e.g.,][]{Hop:98,Row:93}.
The SF history analysis of \citet{Haa:00} provides measurements at
discrete redshifts in addition to the derived evolving LF. These points
cannot be converted directly to the chosen SFR calibration, since this
is luminosity-dependent. In order to be able to show these results, however,
the ratio of each of these points to the measured evolving LF at the
appropriate redshift was noted. These factors were then applied to
the evolving LF after the cosmology and SFR calibration conversions,
and the resulting data points are those given in Table~\ref{datatab}
and shown in Figures~\ref{fig1} and \ref{fig2}. This is a somewhat coarse
way to make the conversion, but it seems unlikely to have introduced an
error of more than about $15\%$ given the locations of these points
with respect to the nearby FIR data points from \citet{Flo:99} both before
and after the conversion.

To convert evolving LFs to the assumed cosmology, in order to compare
with the current data compilation, the following relation is used
\citep[see discussion of Equation~4 from][]{DP:90}:
\begin{equation}
\phi_1(L_1,z)\frac{dV_{c1}}{dz} = \phi_2(L_2,z)\frac{dV_{c2}}{dz}
\end{equation}
where $L_1$ is the luminosity derived from a given flux density
and the luminosity distance corresponding to $z$ in the first
cosmology, and $L_2$ is the luminosity derived from the
same flux density with the luminosity distance for $z$ in the
second cosmology. In other words, $L_2=L_1\times d_{l2}^2/d_{l1}^2$,
with the $d_l$ being the luminosity distances
in the respective cosmologies.

The evolving 1.4\,GHz SF galaxy LFs, after
the cosmology conversion, are converted to SFR densities at each
redshift. This is done simply by integrating the LF at each redshift
after converting luminosity to SFR. As well as the evolving LF of
\citet{Haa:00}, the local 1.4\,GHz LF for SF galaxies from \citet{Con:89}
is also used by invoking pure luminosity evolution, $L\propto(1+z)^Q$,
having a cutoff at $z=2$. Constraints from faint radio source counts
suggest that $2.5 \lesssim Q \lesssim 4.1$ \citep[e.g.,][]{Hop:98}, and
the evolution of the SFR density corresponding to these two extremes
is calculated. The cosmology conversion in both cases is performed
after the application of the appropriate evolutionary factors,
since the derivation of these parameters was performed assuming
an EdS cosmology. The resulting SF histories derived assuming
$Q=2.5$ and $Q=4.1$ (in the EdS cosmology) are shown as dotted
lines in Figures~\ref{fig1}, \ref{fig2} and \ref{fig3}.

\section{Constraints on the LF evolution}
\label{lfconstraint}
The above comparisons illustrate the utility of the SF history to provide
constraints on the form of the LF evolution for SF galaxies.
Earlier studies have used the 1.4\,GHz radio
sources counts with some success to constrain the rate of pure luminosity
evolution ($L\propto(1+z)^Q$) for the SF galaxy population
\cite[e.g.,][]{Row:93,Hop:98,Sey:04}, while assuming no density evolution
($\phi\propto(1+z)^P$, with $P=0$). The constraints
on these forms of evolution from radio source counts suffer from a
degeneracy between luminosity and density evolution \citep{Hop:03b}
in the same fashion as found for the optical source counts
\citep[e.g.,][]{Lin:99}. By itself, the SF history diagram, when used to
constrain the evolution of the LF, suffers from a similar degeneracy.
Together, though, the source counts and the star formation history can
be combined to break this degeneracy, and this is explored here.

The local luminosity function for star forming galaxies of \citet{Sad:02}
was adopted. This was allowed to evolve assuming combinations of density
evolution spanning $-6\le P \le 6$ in steps of 0.31, and luminosity evolution
spanning $-1\le Q \le 7$ in steps of 0.21. In addition a cutoff in luminosity
evolution was imposed above $z=2$ such that for higher redshifts
$L\propto(1+2)^Q$. This is the same cutoff as used in earlier estimates
of pure luminosity evolution \citep{Row:93,Hop:98}. The constraint
derived from the source counts does not turn out to depend strongly on
this cutoff or, indeed, on the assumed cosmology
\citep[this was also recently shown by][]{Sey:04}, although this is not
true for the SF history constraint. For each value of $(P,Q)$ both the source
counts and the star formation history were predicted and
$\chi^2=\Sigma_i (y_{\rm model} - y_i)^2/\sigma_i^2$ calculated.
Here $y_i$ and $\sigma_i$ are either the source count measurements and
uncertainties, respectively, or the SFR density measurements and
uncertainties. The radio source counts used in the $\chi^2$ estimation are
those from the {\em Phoenix Deep Survey\/} (PDS) data of \citet{Hop:03a} for
flux densities $S_{1.4}<5\,$mJy and from the FIRST survey \citep{Whi:97} for
flux densities $3.1<S_{1.4}<502\,$mJy. In addition to the star forming
galaxies, to fit the source counts an active galactic nucleus (AGN) population
is also necessary. The evolving AGN LFs from \citet{DP:90} were adopted, again
as used in earlier studies, with appropriate cosmology conversion.

The SF history measurements used to estimate $\chi^2$,
(where now each $y_i$ and $\sigma_i$ is an SFR density and uncertainty)
are those derived using the common obscuration correction. Only the
independent data points from \citet{Hog:98} from this compilation
were used, and the data point derived from \citet{Con:89} was excluded as its
very small uncertainty otherwise dramatically affects the resulting $\chi^2$.
The GOODS data points at high redshift were also excluded as they are
effectively lower limits, due to the incomplete LF integration, as described
above. The resulting $\chi^2$ arrays were converted to reduced $\chi^2$
values using 58 degrees of freedom (there are 6 parameters in the model,
4 from the LF and the 2 evolutionary parameters) for the source counts
(with 64 measurements) and 48 for the SF history (with 54 measurements).
These reduced $\chi^2$ arrays were then converted to log-likelihoods
using $\log(L)\propto -0.5\chi^2$.

The joint likelihood for each $(P,Q)$ pair (assuming the measurements for
the source counts and the SF history can be treated as independent)
is simply the product of the likelihoods (the sum of the log-likelihoods)
derived from the source counts and the SF history. The assumption of
independence is not trivial, since both the SF history and the source
counts involve integrations over the same luminosity function (weighted,
in the case of the SF history, by the luminosity). For the SF history,
the integration is over luminosity, while for the source counts it is
over redshift. As an analogy, this can be thought of as taking a distribution
of points in flux density/redshift space, and projecting them as distributions
over either redshift or flux density. This would correspond to the
integrations over flux density (or luminosity, for a fixed redshift), or
redshift, respectively. If the distribution of points in flux density/redshift
space is not correlated, the resulting projections are independent, and the
corresponding probability distributions can be assumed to be independent.
Since this is true for deep radio surveys such as the PDS, the assumption of
independence is likely to be reasonable.

\begin{figure*}
\centerline{\rotatebox{0}{\includegraphics[width=9cm]{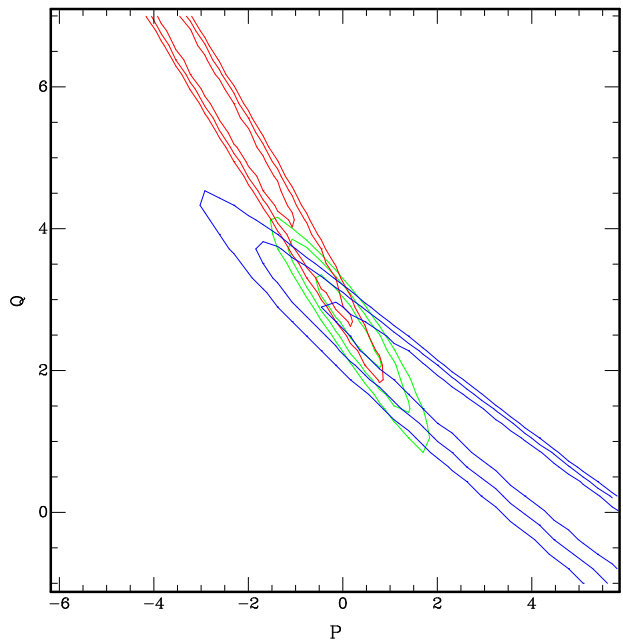}}}
\caption{Probability contours showing $1\,\sigma$, $2\,\sigma$ and
$3\,\sigma$ likelihood regions in the $P-Q$ plane. Blue contours are
the constraint from the sub-mJy radio source counts, red are from the
star formation history diagram, and green are the joint constraint.
 \label{fig4}}
\end{figure*}

Figure~\ref{fig4} shows the 1, 2 and $3\,\sigma$ contours for the constraints
from both the source counts and the SF history, as well as the joint
constraint. These contours are defined by subtracting 0.5, 1.0 and 1.5
respectively from the maximum of $\log(L)$, and the constraints from
the SF history and the source counts seem to be consistent with each
other at about the $2\,\sigma$ level. The maximum likelihood
evolutionary parameters resulting from the joint constraint is
$Q=2.70\pm0.60$, $P=0.15\pm0.60$.

\begin{figure*}
\centerline{\rotatebox{-90}{\includegraphics[width=10cm]{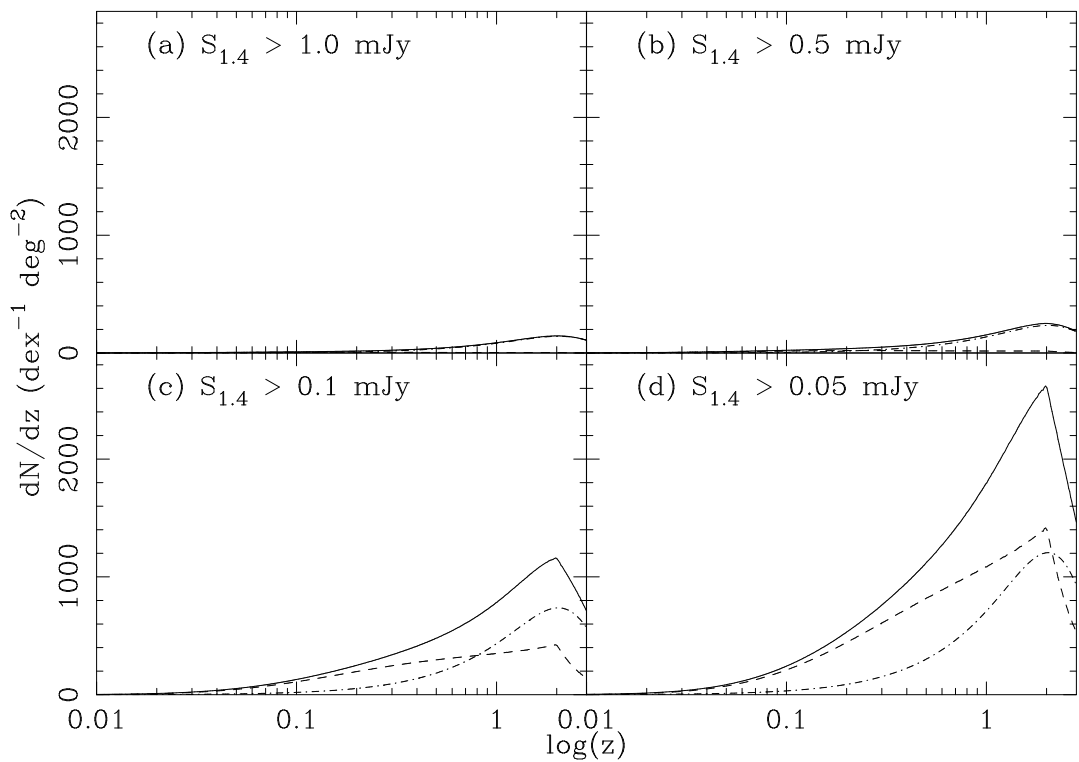}}}
\caption{Redshift distributions predicted from the evolving radio LFs for
four different radio flux density limits. The SF galaxy LF assumes the
best-fitting luminosity and density evolution derived from the previous
Figure ($Q=2.70$, $P=0.15$). In each panel the dot-dashed line is the
contribution from AGN galaxies, the dashed line is that from SF galaxies,
and the solid line is the total galaxy redshift distribution.
 \label{fig5}}
\end{figure*}

To provide some further context for this result, the predicted redshift
distributions for radio sources are shown in Figure~\ref{fig5},
with the SF galaxy contribution derived assuming these rates of evolution.
Four flux density limits are shown, ranging from 1\,mJy, above which
almost all radio sources are AGN dominated, to 0.05\,mJy, the level to which
the 1.4\,GHz source counts are currently robustly measured, and where SF
galaxies dominate the source numbers. It can be seen in Figure~\ref{fig5}(d),
corresponding to the flux density limit relevant for the current analysis,
that SF galaxies dominate the distribution for redshifts below about
$z\approx0.7$, while AGNs contribute a significant proportion only at higher
redshifts. This, along with the other limitations on this analysis at
higher redshift (such as the assumed redshift cutoff in the SF galaxy
evolution and the lack of a large number of SF history estimates),
implies that the current analysis is perhaps less sensitive to the nature
of galaxy evolution at high redshifts, and is constrained most strongly by
the nature of low to moderate redshift SF galaxies. Altering the parameters
of the assumed AGN luminosity function within a small range (constrained
to retain the consistency with the observed radio source counts), for example,
has the effect of modestly changing the shape of the predicted redshift
distribution for the AGNs. Since these predominantly lie at high redshift
and bright flux densities, the radio source count constraints for the
evolution of the SF galaxy population are not significantly affected.

The assumed redshift cutoff in the SF galaxy evolution above $z=2$ can be
seen as a sharp turnover in the SF galaxy numbers at this redshift, most
noticable in Figure~\ref{fig5}(c) and \ref{fig5}(d). This type of artifact is
unlikely to appear in more realistic models of galaxy evolution and reflects
the illustrative nature of the current investigation, which obviously leaves
some room for refinement. The other aspect that must be taken
into account in any interpretation of the distributions in this Figure
is the effect of the optical counterpart brightness in constraining any
observational redshift distributions that may be obtained. Models of
bivariate radio-optical LFs and corresponding forms of evolution have
been used to predict such distributions, \citep[e.g.,][]{Hop:98b,Hop:99},
although such analysis is beyond the scope of the present work.

The likelihood contours for the SF history and source count constraints seen
in Figure~\ref{fig4} show fairly well defined slopes, corresponding to
degeneracies in $(P,Q)$. The reasons for these degeneracies can be
understood in a little more detail by considering the dependence on
$P$ and $Q$ of the calculations used to produce the SF history and the
source counts. The source counts are considered here first, and can
be written as:
\begin{equation}
S N(S) \propto \int dz \frac{dV_c}{dz} L\phi(L)
\end{equation}
where $L=4\pi S [(1+z)D_c(z)]^2$, and $V_c=4\pi D_c^3/3$, and the subscript
$c$ indicates the comoving distance or volume. The
adopted forms for the luminosity and density evolution, $L^*(z)=L^*_0(1+z)^Q$
and $\phi^*(z)=\phi^*_0(1+z)^P$, mean that $L\phi(L)$
can be written as $(1+z)^P \phi^*_0 f(y)$, where $y=L/L^*(z)$. Now, the
comoving distance can be approximated as $D_c(z) = D_{EdS}(z) (1+z)^\gamma$
with the Einstein-de Sitter comoving distance
$D_{EdS}(z) = 2(c/H)(1 - 1/\sqrt{1+z})$, and $\gamma$ being a function of
$\Omega$ and $\Lambda$. The expression $y$ is then a complicated function
of $(1+z)$. Transforming the integral over $z$ into one over $y$ gives
\begin{equation}
 S N(S) \propto \int dy {{dz}\over{dy}} {{dD_c}\over{dz}}
                       {{dV_c}\over{dD_c}}\, (1+z)^P f(y).
\end{equation}
Then, since
\begin{eqnarray}
 y &=& {{4\pi S}\over{L^*_0}} {{(1+z)^2\, D_c^2(z)}\over{(1+z)^Q}}, \nonumber \\
 {{dy}\over{dz}} &=& {{y}\over{1+z}}
                 \left({{1}\over{\sqrt{1+z}-1}} + 2\gamma + 2-Q\right), \nonumber \\
 {{dD_c}\over{dz}} &=& {{D_c(z)}\over{1+z}}
     {{1 + 2\gamma (\sqrt{1 + z} - 1 )}\over{2 (\sqrt{1+z} - 1)}}, \qquad {\rm and} \nonumber \\
 {{dV_c}\over{dD_c}} &=& 4\pi D_c^2, \nonumber
\end{eqnarray}
the expression for the source counts becomes
\begin{eqnarray}
 S N(S) &\propto& \int {{dy}\over{2y}}\,f(y)
 \left[{{1 + 2\gamma (\sqrt{1 + z} - 1 )}\over{1 + (2\gamma + 2-Q)(\sqrt{1+z}-1)}}
 \right] \nonumber \\
 & & \times (1+z)^P\, D_c^3.
\end{eqnarray}
This is a relation that can now be used to estimate the form of the
degeneracy between $P$ and $Q$, given the observed source counts.
The observed source count slope is close to Euclidean for small $z$,
{\em i.e.}, the differential source count $N(S) \propto S^{-5/2}$, or
$S N(S) \propto S^{-3/2}$, and this should be independent of $z$ for small
$z$. Now $D_c^3 \propto (y/S)^{3/2}(1+z)^{3Q/2 - 3}$, and
at small $z$ the term above in square brackets is
$(1+\gamma z)/(1 + (\gamma + 1 - Q/2)z)\approx (1+z)^{Q/2-1}$, so
\begin{equation}
 S N(S) \propto \int {{dy}\over{2y}}\, \left({{y}\over{S}}\right)^{3/2}\,f(y)
 (1+z)^{P + 3Q/2 - 3 + Q/2 - 1},
\end{equation}
at least in the regime in which the counts are dominated by objects
at relatively small redshifts. Since there should be no dependence on
$(1+z)$ in this regime, it might thus be expected that the constraint
from the number counts scales as
\begin{equation}
P + 2Q \approx 4.
\end{equation}
This is independent of $\gamma$ since at small $z$ all cosmologies
have similar $D_c(z)$ relations. The degeneracy indicated by the actual
numerical estimates giving the likelihood contours in Figure~\ref{fig4} is
$P + 2Q \approx 5.4$. The difference from the scaling derived above
is because, as might be expected, the contribution to the source counts
from higher redshifts is not negligible.

The evolution of the SF history is somewhat simpler to understand.
Since most SFR indicators are proportional to luminosity, $\dot{\rho}_*(z)$
is just the evolution of the luminosity density,
\begin{eqnarray}
 \dot{\rho}_*(z)&\propto& \int dL\,\phi(L|z)\,L \nonumber \\
             &\propto& L^*(z) \int dL\,\phi(L|z)\,{{L}\over{L^*(z)}} \nonumber \\
             &\propto& L^*_0\,(1+z)^Q\, \phi^*_0\,(1+z)^P\, \int dy\,f(y).
\end{eqnarray}
The observed slope of $\log{\dot{\rho}_*(z)}$ with $\log(1+z)$ thus
constrains the value of the sum $(P+Q)$. This slope has been seen to be
about 3 (up to $z=1$), so $P+Q\approx 3$, which is in excellent agreement with
the degeneracy in $(P,Q)$ seen in Figure~\ref{fig4} for the SF history
likelihood contours.

This analysis suggests that there are understandable reasons for the
degeneracies seen between $P$ and $Q$, and the measurements are consistent
with what might be expected from this analysis. The form of the degeneracies
are fortunately different for the source counts and the SF history, and
the degeneracies can be broken by combining these constraints. More detailed
investigation along these lines might usefully identify additional observables
that could similarly be used to break such degeneracies, yet further
constraining the form of evolution of the star forming galaxy luminosity
function.

\section{Discussion}
\label{disc}
\subsection{Limitations of this analysis}
\label{disc1}
There appears to be a large discrepancy (factors of two) in the local
SFR density derived from integrating various local radio LFs. When the
actual LFs are compared, however, the measurements are highly consistent
(compare, e.g., the LF and parameterisations of \citet{Sad:02} and
\citet{Mac:00}). The discrepancy here occurs because of the different
LF parameters derived, (even for the same functional form), which when
integrated over the complete luminosity range give different results. This is
in part a contribution of the poorly constrained tails of the LFs, but there
is also a contribution from the different values for $L^*$ which translate
to different locations for the peak in the luminosity density as a function of
luminosity, and hence to the derived total SFR densities. It is known that
LF parameters are not independent, and the effect of this is that the
uncertainty in the LF integral requires the incorporation of the LF
parameter error ellipse, in addition to the Poisson uncertainties typically
quoted. The result is that at least some of the scatter in the SF history
diagram will be due to different choices for LF parameterisation (not only
for radio LFs, but also at other wavelengths).

This systematic uncertainty is clearly not accounted for in the published
uncertainties for the SFR density measurements. These typically reflect
only assumed Poisson counting uncertainties, while neglecting other systematics
that may be important for any particular survey, typically because these
effects are difficult to quantify. As a result the quoted uncertainties
for most data points are likely to be lower limits to the true uncertainties.
This has not been accounted for in the $\chi^2$ calculations used above to
constrain the SF galaxy LF evolution. One way to minimise any bias
introduced by this issue might be to bin the measured SFR density
points in redshift, summing in quadrature their published uncertainties.
By combining adjacent data points as an average or median, the most
extreme outlying points as well as those with the least representative
uncertainties, will have their disproportionate effects on the $\chi^2$
statistic minimised. This has been explored using several different
binnings of the data, ranging from including as few as 5 points per bin
to as many as 8 points per bin (coarser binning leaves the problem
unconstrained as there are then fewer data points than free parameters,
while finer binning eliminates the advantage of this method as there
are then too few points per bin to smooth out the more extreme measurements).
The joint constraint with the source counts when using this method
is changed only marginally, although the likelihood peak in the
constraint from the SF history alone moves along the line of degeneracy
to smaller $Q$ and larger $P$ values. This suggests that the apparent
marginal consistency (at the $2\,\sigma$ level) between the SF history
and source count datasets separately is likely to be an artifact of
a few measurements contributing disproportionately to the $\chi^2$
estimate for the former, and the true consistency could be much better.

The preferred value for the luminosity evolution, $Q=2.70\pm0.60$, is highly
consistent with a recent estimate from models incorporating luminosity
evolution only (no density evolution) when constrained by the 1.4\,GHz
source counts \citep{Sey:04}. It is worth noting, though, that is somewhat
lower than the fitted slope to the SF history data points for
$z<1$, $\beta=3.10\pm0.25$. This cannot be completely explained by the
small contribution from the density evolution ($P=0.15\pm0.60$), since
even together these effects produce a slightly flatter slope compared with
$\beta$. This is a result both of including points at $z>1$ in the
$\chi^2$ calculation as well as the effect of the constraint from the
source counts (the constraint from the SF history alone seems to favour
somewhat higher values of $Q$ with negative values for $P$). Regarding
the choice of local SF galaxy LF for the investigation of the evolutionary
parameters, choosing a different LF from that of \citet{Sad:02} would produce
slightly different results for the best-fitting values of $(P,Q)$. The LF
from \citet{Con:02} (producing one of the lowest values for the local SFR
density) would favour somewhat higher rates of luminosity or density
evolution, and a correspondingly steeper slope in the global SF history
diagram. When the constraints from the source counts are incorporated, though,
the resulting preferred values still do not change much from those
derived above.

Apart from the factors of two uncertainty in the local estimate for
the SFR density, the other most strikingly outlying points include
the lowest redshift UV-derived points \citep{Sul:00,Trey:98}, and
the [O{\sc ii}] data from \citet{Hog:98}. The latter is explained
by that author simply as the effect of the small numbers contributing
to each redshift bin, and the choice of bin location (prompting the
use of the overlapping bins chosen to emphasise the extent of this effect).
The UV-derived estimates at $z=0.15$ appear to be high compared with
other estimates from H$\alpha$, [O{\sc ii}] or radio luminosities at
similar redshifts. It is likely that this is actually a result of
an {\em overestimated\/} obscuration correction, both with the
common obscuration and the SFR-dependent obscuration methods used,
which can be seen to produce similar corrections. The underlying reason for
this is a combination of lower luminosity systems having lower obscurations,
and low-redshift UV-selected samples being dominated primarily by such
relatively low luminosity systems. This is supported by the fact that the SFR
density derived for the samples of \citet{Tres:98} and \citet{Trey:98} using
the SFR dependent obscuration gives a {\em lower\/} value than that using
the common obscuration correction. Further, as can be seen from
Figure~1 of \citet{Sul:00}, for example, although the mean obscuration is
close to $A_V=1\,$mag for this sample, the median is somewhat lower (probably
closer to $A_V=0.8\,$mag). The effective obscuration correction for this sample, in terms of a common obscuration, should thus
be somewhat lower than assumed. For the luminosity-dependent obscuration,
similarly, the curve from \citet{Hop:01} is again less appropriate, having
been derived from a sample containing more heavily obscured systems
at all luminosities than are likely to be found in the local UV-selected
samples. When these effects are accounted for, the UV-derived points at
$z=0.15$ will be moved lower, more consistent with the rest of the
compilation around this redshift.

\subsection{The shape of the $z>3$ SF history}
\label{disc2}
Recent investigations of $z\approx6$ $i'$-band dropouts have suggested that
the SF history is steadily dropping from $z=3$ to $z\approx6$
\citep{Bou:03,Fon:03,Sta:04}. These analyses compare the comoving SFR density
for objects with high SFRs \citep[$\gtrsim 15\,M_{\odot}$yr$^{-1}$,][]{Sta:04},
and find a decrease in this SFR density of almost an order of magnitude
from $z=3$ to $z\approx6$. The reason for limiting the analysis to objects
with high SFRs is to avoid the extrapolations made in assuming a particular
faint end slope ($\alpha$) for the LF. If the faint end slope is assumed
to be $\alpha=-1.6$ \citep{Yan:03,Bou:03}, and the LF integrated over all
luminosities, the derived SFR density is a factor of about four greater
than for just the high SFR objects alone \citep{Bou:03}. The value established
assuming this faint end slope by \citet{Bou:03} (after conversion to the UV
SFR calibration adopted here) is $0.04\,M_{\odot}$yr$^{-1}$. If obscuration
corrections are then made (assuming the common obscuration correction used
herein), the SFR density is increased by another factor of about 3.4, to
$0.14\,M_{\odot}$yr$^{-1}$ at $z\approx6$, highly consistent with the other
estimates shown in Figure~\ref{fig1}. It should be emphasised that the
estimated $\approx 40\%$ decrease in comoving SFR density from $z=3$ to
$z\approx6$ claimed by \citet{Bou:03} lies entirely within the scatter of the
measurement compilation, and associated uncertainties, over this redshift
range. The much lower $z\approx 6$ SFR densities derived by \citet{Fon:03} and
\citet{Sta:04}, indicating an order of magnitude drop from $z=3$, are less
consistent, being significantly lower than the estimates of \citet{Bou:03}.
These apparent inconsistencies are related to the small numbers of objects
in the samples (at most about 30, with as few as 6 galaxies contributing
to some analyses), the assumptions involved in correcting for incompleteness,
and the uncertainty in the faint end slope of the luminosity function.
Requiring an absolute magnitude limited sample, for example, seems to imply
constant SFR densities up to $z\approx 6$ \citep{Fon:03}. There is
evidence, too, for faint end LF slopes as steep as $\alpha=-1.6$ or
steeper \citep{Yan:03} at these redshifts, which would also support relatively
little evolution.

Given these uncertainties, it is clear that larger samples
are needed at these high redshifts, probing to fainter
luminosities to robustly determine the faint end slope of the LF,
before it is possible to reliably constrain the shape of the
SFR density evolution beyond $z\approx3$.

\section{Summary}
\label{summ}
An extensive compilation from the literature of SFR density measurements
as a function of redshift has been presented. These data have been
converted to a consistent set of SFR calibrations, common cosmology,
and consistent dust obscuration corrections where necessary. The
compilation thus produced gives a highly consistent description of
the evolution in galaxy SFR density with redshift, constrained to
within factors of about three at most redshifts up to $z\approx6$.
In producing this compilation, different assumptions regarding
obscuration have been explored. For those measurements requiring obscuration
corrections the assumptions used were those deemed most likely to be
appropriate. Alternative assumptions regarding the obscuration corrections,
however, will act to {\em increase\/} the SFR densities derived compared
with those presented here (with the possible exception of the poorly
known obscuration properties of high redshift systems, $z\gtrsim1$).

This rich compilation of SFR density evolution provides a robust constraint
for many investigations of galaxy evolution, and this has been illustrated
by constraining the evolution of the SF galaxy LF. In combination with
constraints from the radio source counts, the SF galaxy LF is found to
evolve with luminosity and density evolutionary parameters $Q=2.70\pm0.60$
and $P=0.15\pm0.60$ respectively. An analysis has been performed of the
degeneracies between $P$ and $Q$ seen in the separate constraints from
the source counts and the SF history. This suggests that the origin of
these degeneracies is reasonably well understood, and this methodology
may be a useful tool for identifying additional observables that could be
used as additional constraints for breaking such degeneracies.

While every effort has been made to be as thorough as possible in the
current compilation, it is possible that some datasets or measurements may
have been inadvertently omitted. It seems unlikely, though, that such
omissions will significantly alter the above conclusions. At the same time,
it is the author's hope that a complete and expansive review of this topic
(more so than possible in the present case) will be able to do justice
to the wealth of current and ongoing measurements contributing to our
understanding of this aspect of galaxy evolution.

\acknowledgements
The referee is thanked for several insightful comments that
have improved this compilation and analysis.
I gratefully thank Ravi Sheth for contributing the detailed analysis of the
evolutionary dependencies discussed in \S\,\ref{lfconstraint},
and for additional helpful discussion.
I also thank Andrew Connolly, Simon Krughoff, Chris Miller, and
Ryan Scranton for providing detailed and very helpful discussion on many
aspects of this work.
I acknowledge support provided by the National Aeronautics and Space
Administration (NASA) through Hubble Fellowship grant
HST-HF-01140.01-A awarded by the Space Telescope Science Institute (STScI).
STScI is operated by the Association of Universities for Research in
Astronomy, Inc., under NASA contract NAS5-26555.

\begin{deluxetable}{ll}
\tablecaption{SFR calibrations. SFR is in units of $M_{\odot}\,$yr$^{-1}$.
1.4\,GHz and UV luminosities are in units of W\,Hz$^{-1}$. FIR and H$\alpha$
luminosities are in units of W.
 \label{sfrcal}}
\tablehead{
\colhead{Wavelength} & \colhead{Calibration}
}
\startdata
1.4\,GHz & ${\rm SFR} = \left\{\begin{array}{l} L_{1.4}/1.81\times10^{21} \hspace{49.5mm} L_{1.4}>6.4\times10^{21}\,{\rm W\,Hz}^{-1} \\ L_{1.4}/[1.81\times10^{21}(0.1+0.9(L_{1.4}/6.4\times10^{21})^{0.3}] \hspace{3mm} L_{1.4}<6.4\times10^{21}\,{\rm W\,Hz}^{-1} \end{array}\right.$ \\
FIR & ${\rm SFR} = L_{\rm FIR}/2.22\times10^{36}$ \\
H$\alpha$ & ${\rm SFR} = L_{\rm H\alpha}/1.26\times10^{34}$ \\
UV & ${\rm SFR} = L_{\rm UV}/7.14\times10^{20}$ \\
\enddata
\end{deluxetable}

\begin{deluxetable}{lccccccc}
\tablecaption{Measurements of SFR density, $\dot{\rho}_*$, in units of
$M_{\odot}\,$yr$^{-1}$\,Mpc$^{-3}$, as a function of redshift.
 \label{datatab}}
\tablehead{
\colhead{Reference} & \colhead{Estimator} & \colhead{Redshift} & \colhead{$C_1$\tablenotemark{a}} & \colhead{$\log(\dot{\rho}_{*{\rm com}})\tablenotemark{b}$} & \colhead{$C_2$\tablenotemark{c}} & \colhead{$\log(\dot{\rho}_{*{\rm sfrd}})\tablenotemark{d}$} & \colhead{$C_3$\tablenotemark{e}}
}
\startdata
\citet{Gia:04} & 1500\,\AA\ &  $3.780\pm0.340$ & 1.0 & $-0.772\pm0.060$ & \ldots & & \\
 & &  $4.920\pm0.330$ & 1.0 & $-0.963\pm0.140$ & \ldots & & \\
 & &  $5.740\pm0.360$ & 1.0 & $-0.923\pm0.190$ & \ldots & & \\
\citet{Wil:02} & 2500\,\AA\ &  $0.350\pm0.150$ & 1.0 & $-1.449\pm0.078$ & 2.53 & $-1.366\pm0.097$ & 3.06 \\
 & &  $0.800\pm0.200$ & 1.0 & $-1.183\pm0.080$ & 2.53 & $-1.129\pm0.043$ & 2.87 \\
 & &  $1.350\pm0.250$ & 1.0 & $-1.034\pm0.143$ & 2.53 & $-1.016\pm0.093$ & 2.64 \\
\citet{Mas:01} & 1500\,\AA\ &  $1.500\pm0.500$ & 0.822 & $-0.709\pm0.200$ & \ldots & $-0.709\pm0.200$ & \ldots \\
 & &  $2.750\pm0.750$ & 0.784 & $-0.512\pm0.200$ & \ldots & $-0.512\pm0.200$ & \ldots \\
 & &  $4.000\pm0.500$ & 0.774 & $-0.886\pm0.250$ & \ldots & $-0.886\pm0.250$ & \ldots \\
\citet{Sul:00} & 2000\,\AA\ &  $0.150\pm0.150$ & 0.611 & $-1.403\pm0.050$ & 2.86 & $-1.402\pm0.050$ & 2.87 \\
\citet{Ste:99} & 1700\,\AA\ &  $3.040\pm0.250$ & 0.389 & $-0.795\pm0.050$ & 3.15 & $-0.585\pm0.050$ & 5.11 \\
 & &  $4.130\pm0.300$ & 0.387 & $-0.905\pm0.100$ & 3.15 & $-0.693\pm0.100$ & 5.13 \\
\citet{Cow:99} & 2000\,\AA\ &  $0.700\pm0.200$ & 0.717 & $-1.318\pm0.104$ & 2.86 & $-1.316\pm0.052$ & 2.88 \\
 & &  $1.250\pm0.250$ & 0.646 & $-1.186\pm0.117$ & 2.86 & $-1.057\pm0.060$ & 3.86 \\
\citet{Trey:98} & 2000\,\AA\ &  $0.150\pm0.150$ & 0.611 & $-1.369\pm0.150$ & 2.86 & $-1.439\pm0.150$ & 2.44 \\
\citet{Con:97} & 2800\,\AA\ &  $0.750\pm0.250$ & 0.917 & $-0.992\pm0.150$ & 2.37 & $-0.887\pm0.150$ & 3.01 \\
 & &  $1.250\pm0.250$ & 0.841 & $-0.864\pm0.150$ & 2.37 & $-0.679\pm0.150$ & 3.63 \\
 & &  $1.750\pm0.250$ & 0.812 & $-0.974\pm0.150$ & 2.37 & $-0.717\pm0.150$ & 4.29 \\
\citet{Lil:96} & 2800\,\AA\ &  $0.350\pm0.150$ & 1.07 & $-1.539\pm0.070$ & 2.37 & & \\
 & &  $0.625\pm0.125$ & 0.953 & $-1.266\pm0.080$ & 2.37 & & \\
 & &  $0.875\pm0.125$ & 0.892 & $-0.979\pm0.150$ & 2.37 & & \\
\citet{Mad:96} & 1600\,\AA\ &  $2.750\pm0.750$ & 0.784 & $>-1.217$ & 3.27 & & \\
 & &  $4.000\pm0.500$ & 0.774 & $>-1.723$ & 3.27 & & \\
\citet{Tep:03} & [O{\sc ii}] &  $0.900\pm0.500$ & 1.0 & $-1.005\pm0.110$ & 2.51 & $-0.902\pm0.110$ & 3.18 \\
\citet{Gal:02} & [O{\sc ii}] &  $0.025\pm0.025$ & 0.677 & $-1.913\pm0.150$ & \ldots & $-1.913\pm0.150$ & \ldots \\
\citet{Hog:98} & [O{\sc ii}] &  $0.200\pm0.100$ & 0.588 & $-1.640^{+0.163}_{-0.118}$ & 2.51\tablenotemark{f} & & \\
 & &  $0.300\pm0.100$ & 0.551 & $-1.712^{+0.101}_{-0.094}$ & 2.51\tablenotemark{f} & & \\
 & &  $0.400\pm0.100$ & 0.522 & $-1.067^{+0.105}_{-0.085}$ & 2.51\tablenotemark{f} & & \\
 & &  $0.500\pm0.100$ & 0.499 & $-0.826^{+0.075}_{-0.067}$ & 2.51\tablenotemark{f} & & \\
 & &  $0.600\pm0.100$ & 0.480 & $-1.003^{+0.072}_{-0.062}$ & 2.51\tablenotemark{f} & & \\
 & &  $0.700\pm0.100$ & 0.466 & $-1.094^{+0.088}_{-0.073}$ & 2.51\tablenotemark{f} & & \\
 & &  $0.800\pm0.100$ & 0.454 & $-1.076^{+0.081}_{-0.071}$ & 2.51\tablenotemark{f} & & \\
 & &  $0.900\pm0.100$ & 0.444 & $-0.982^{+0.090}_{-0.074}$ & 2.51\tablenotemark{f} & & \\
 & &  $1.000\pm0.100$ & 0.436 & $-0.783^{+0.107}_{-0.087}$ & 2.51\tablenotemark{f} & & \\
 & &  $1.100\pm0.100$ & 0.429 & $-0.893^{+0.199}_{-0.136}$ & 2.51\tablenotemark{f} & & \\
 & &  $1.200\pm0.100$ & 0.423 & $-0.919^{+0.301}_{-0.176}$ & 2.51\tablenotemark{f} & & \\
\citet{Ham:97} & [O{\sc ii}] &  $0.375\pm0.125$ & 1.06 & $-1.705^{+0.070}_{-0.080}$ & 2.51\tablenotemark{f} & & \\
 & &  $0.625\pm0.125$ & 0.953 & $-1.226^{+0.110}_{-0.150}$ & 2.51\tablenotemark{f} & & \\
 & &  $0.875\pm0.125$ & 0.892 & $-0.855^{+0.200}_{-0.380}$ & 2.51\tablenotemark{f} & & \\
\citet{Pet:98} & H$\beta$ &  $2.750\pm0.750$ & 0.784 & $-0.557\pm0.150$ & 3.71\tablenotemark{g} & & \\
\citet{Per:03} & H$\alpha$ &  $0.025\pm0.025$ & 1.0 & $-1.604\pm0.110$ & \ldots & $-1.604\pm0.110$ & \ldots \\
\citet{Tres:02} & H$\alpha$ &  $0.700^{+0.400}_{-0.200}$ & 0.933 & $-0.931\pm0.110$ & 2.51 & $-0.933\pm0.110$ & 2.50 \\
\citet{Moo:00} & H$\alpha$ &  $2.200\pm0.050$ & 0.790 & $-0.576\pm0.120$ & 2.51 & $-0.426\pm0.120$ & 3.54 \\
\citet{Hop:00} & H$\alpha$ &  $1.250\pm0.550$ & 0.561 & $-0.629\pm0.026$ & 2.51 & $-0.588\pm0.064$ & 2.76 \\
\citet{Sul:00} & H$\alpha$ &  $0.150\pm0.150$ & 0.611 & $-1.820\pm0.060$ & \ldots & $-1.820\pm0.060$ & \ldots \\
\citet{Gla:99} & H$\alpha$ &  $0.900\pm0.100$ & 0.89 & $-0.972^{+0.150}_{-0.140}$ & \ldots & & \\
\citet{Yan:99} & H$\alpha$ &  $1.300\pm0.600$ & 0.837 & $-0.553\pm0.120$ & 2.51 & $-0.412\pm0.120$ & 2.41 \\
\citet{Tres:98} & H$\alpha$ &  $0.200\pm0.100$ & 1.17 & $-1.489\pm0.060$ & 2.51 & $-1.571\pm0.060$ & 2.09 \\
\citet{Gal:95} & H$\alpha$ &  $0.022\pm0.022$ & 1.37 & $-1.900\pm0.200$ & \ldots & $-1.900\pm0.200$ & \ldots \\
\citet{Flo:99} & $15\,\mu$m &  $0.350\pm0.150$ & 1.07 & $-1.438\pm0.270$ & \ldots & $-1.438\pm0.270$ & \ldots \\
 & &  $0.625\pm0.125$ & 0.953 & $-1.169\pm0.250$ & \ldots & $-1.169\pm0.250$ & \ldots \\
 & &  $0.875\pm0.125$ & 0.892 & $-0.874\pm0.260$ & \ldots & $-0.874\pm0.260$ & \ldots \\
\citet{Bar:00} & $850\,\mu$m &  $2.000\pm1.000$ & 0.615 & $-0.831^{+0.220}_{-0.230}$ & \ldots & $-0.831^{+0.220}_{-0.230}$ & \ldots \\
 & &  $4.500\pm1.500$ & 0.594 & $-0.721^{+0.360}_{-0.430}$ & \ldots & $-0.721^{+0.360}_{-0.430}$ & \ldots \\
\citet{Hug:98} & $850\,\mu$m &  $3.000\pm1.000$ & 0.39 & $>-1.087\pm0.170$ & \ldots & & \\
\citet{Con:02} & 1.4\,GHz &  $0.005\pm0.005$ & 1.0 & $-1.964\pm0.030$ & \ldots & $-1.964\pm0.030$ & \ldots \\
\citet{Sad:02} & 1.4\,GHz &  $0.080\pm0.080$ & 1.30 & $-1.728^{+0.080}_{-0.090}$ & \ldots & $-1.728^{+0.080}_{-0.090}$ & \ldots \\
\citet{Ser:02} & 1.4\,GHz &  $0.010\pm0.010$ & 1.39 & $-1.753^{+0.080}_{-0.100}$ & \ldots & $-1.753^{+0.080}_{-0.100}$ & \ldots \\
Machalski \& & 1.4\,GHz & $0.070\pm0.070$ & 1.31 & $-1.920\pm0.100$ & \ldots & $-1.920\pm0.100$ & \ldots \\
\hspace{3mm}Godlowski (2000) & & & & & & & \\
\citet{Haa:00} & 1.4\,GHz &  $0.280^{+0.121}_{-0.270}$ & 0.60 & $-1.389^{+0.140}_{-0.210}$ & \ldots & $-1.389^{+0.140}_{-0.210}$ & \ldots \\
 & &  $0.460^{+0.058}_{-0.050}$ & 0.52 & $-1.176^{+0.140}_{-0.200}$ & \ldots & $-1.176^{+0.140}_{-0.200}$ & \ldots \\
 & &  $0.600^{+0.098}_{-0.052}$ & 0.48 & $-1.117^{+0.160}_{-0.260}$ & \ldots & $-1.117^{+0.160}_{-0.260}$ & \ldots \\
 & &  $0.810^{+0.074}_{-0.086}$ & 0.44 & $-0.881^{+0.130}_{-0.180}$ & \ldots & $-0.881^{+0.130}_{-0.180}$ & \ldots \\
 & &  $1.600^{+2.820}_{-0.640}$ & 0.40 & $-0.785^{+0.120}_{-0.180}$ & \ldots & $-0.785^{+0.120}_{-0.180}$ & \ldots \\
\citet{Con:89} & 1.4\,GHz &  $0.005\pm0.005$ & 1.55 & $-1.679\pm0.001$ & \ldots & $-1.679\pm0.001$ & \ldots \\
\citet{Age:03} & X-ray &  $0.24^{+0.06}_{-0.24}$ & 0.88 & $-1.417^{+0.219}_{-0.182}$ & \ldots & $-1.417^{+0.219}_{-0.182}$ & \ldots \\
 & ($0.5-2\,$keV) &  &  & \\
\enddata
\tablenotetext{a}{The factor effectively used in converting $\dot{\rho}_*$ from
the cosmology assumed in the original reference to that assumed here. In cases
where an LF has been integrated this factor is the appropriate combination of
the conversion factors applied separately to $L^*$ and $\phi^*$.}
\tablenotetext{b}{$\dot{\rho}_*$ calculated using the common obscuration
correction.}
\tablenotetext{c}{The factor corresponding to the common obscuration
correction applied in calculating $\dot{\rho}_{*{\rm com}}$. An ellipsis means
that an obscuration correction applied in the original reference was
retained, or that no obscuration correction is necessary.}
\tablenotetext{d}{$\dot{\rho}_*$ calculated using the SFR-dependent
obscuration correction. No entry means that no LF parameters were available
for use in applying an SFR-dependent correction. (Lower limits, too, are 
not transferred to this column.)}
\tablenotetext{e}{The effective factor corresponding to the SFR-dependent
obscuration correction applied in calculating $\dot{\rho}_{*{\rm sfrd}}$. It
should be emphasised that this is merely the inferred effective correction,
as the actual correction is done through an integral over the luminosity
function, while applying corrections varying as a function of luminosity.
An ellipsis means that an obscuration correction applied in the original
reference was retained, or that no obscuration correction is necessary.
No entry means no $\dot{\rho}_{*{\rm sfrd}}$ was available.}
\tablenotetext{f}{Obscuration correction valid at the wavelength of H$\alpha$,
since the [O{\sc ii}] LF is converted to an H$\alpha$ LF before obscuration
correction.}
\tablenotetext{g}{Obscuration correction valid at the wavelength of H$\beta$
corresponding to a 1\,mag correction at the wavelength of H$\alpha$.}
\end{deluxetable}

\begin{deluxetable}{lcccccc}
\tablecaption{Luminosity function parameters, converted to the currently
assumed cosmology.
 \label{lfparams}}
\tablehead{
\colhead{Reference} & \colhead{$z$} & \colhead{$\log(L^*)$} & \colhead{$\phi^*$} & \colhead{$\alpha$} & \colhead{$\sigma$} & Note
}
\startdata
\citet{Wil:02} & 0.35 & 21.182, 21.806 & $5.52$, $1.06$ & $-1.0$, $-1.5$ & \ldots & \\
 & 0.80 & 21.286, 21.474 & $7.96$, $4.22$ & $-1.0$, $-1.5$ & \ldots & \\
 & 1.35 & 21.198, 21.290 & $11.9$, $10.5$ & $-1.0$, $-1.5$ & \ldots & \\
\citet{Sul:00} & 0.15 & 21.342 & $2.48$ & $-1.51$ & \ldots & \\
\citet{Ste:99} & 3.04 & 22.072 & $1.39$ & $-1.6$ & \ldots & \\
 & 4.13 & 22.076 & $1.07$ & $-1.6$ & \ldots & \\
\citet{Cow:99} & 0.70 & 21.05, 21.17 & $8.4$, $5.8$ & $-1.0$, $-1.5$ & \ldots & \\
 & 1.25 & 21.48, 21.56 & $4.1$, $3.3$ & $-1.0$, $-1.5$ & \ldots & \\
\citet{Trey:98} & 0.15 & 21.285 & $2.37$ & $-1.62$ & \ldots & \\
\citet{Con:97} & 0.75 & 21.65 & $5.30$ & $-1.3$ & \ldots & \\
 & 1.25 & 22.01 & $3.10$ & $-1.3$ & \ldots & \\
 & 1.75 & 22.35 & $1.10$ & $-1.3$ & \ldots & \\
\citet{Tep:03} & 0.90 & 35.60 & $0.90$ & $-1.35$ & \ldots & 1. \\
\citet{Gal:02} & 0.025 & 36.33 & $0.0636$ & $-1.17$ & \ldots & 1., 2. \\
\citet{Per:03} & 0.025 & 35.43 & $1.00$ & $-1.2$ & \ldots & 2. \\
\citet{Tres:02} & 0.7 & 34.97 & $4.80$ & $-1.31$ & \ldots & \\
\citet{Moo:00} & 2.2 & 35.88 & $1.27$ & $-1.35$ & \ldots & 3. \\
\citet{Hop:00} & 1.25 & 35.87, 36.34 & $0.766$, $0.088$ & $-1.60$, $-1.86$ & \ldots & \\
\citet{Sul:00} & 0.15 & 35.42 & $0.313$ & $-1.62$ & \ldots & 2. \\
\citet{Yan:99} & 1.3 & 35.83 & $1.50$ & $-1.35$ & \ldots & \\
\citet{Tres:98} & 0.20 & 34.61 & $2.89$ & $-1.35$ & \ldots & \\
\citet{Gal:95} & 0.0225 & 34.87 & $1.65$ & $-1.3$ & \ldots & 2. \\
\citet{Sad:02} & 0.08 & 19.29 & $22.9$ & $0.84$ & 0.94 & 4. \\
\citet{Ser:02} & 0.01 & 22.16 & $1.32$ & $-1.29$ & \ldots & \\
\citet{Mac:00} & 0.07 & 21.468 & $2.38$ & $1.22$ & 0.61 & 4. \\
\enddata
\tablecomments{Units of $L^*$ are W for H$\alpha$ LFs, or W\,Hz$^{-1}$ for
UV or 1.4\,GHz LFs. Units of $\phi^*$ are $10^{-3}$\,Mpc$^{-3}$. The 1.4\,GHz
LFs from \citet{Con:02} and \citet{Con:89} use a parameterisation
corresponding to ``hyperbolic" visibility functions. These are parameterised
using $Y=3.06$, $B=1.9$, $X=22.35$, $W=0.67$ \citep[for][]{Con:02} and
$Y=2.88$, $B=1.5$, $X=22.108$, $W=0.667$ \citep[for][]{Con:89}, after
conversion to the currently assumed cosmology.\\
1. Effective H$\alpha$ LF inferred from observed [O{\sc ii}] LF.\\
2. Obscuration corrected by original authors using measured values of
obscuration for individual objects.\\
3. This LF was not fitted to the observations. Rather the observations at
$z=2.2$ appear consistent with the $z=1.3$ LF of \citet{Yan:99}, in the
cosmologies assumed by those authors, so that LF is adopted for this
higher redshift. Note that the cosmology conversions at the different
redshifts give rise to different $L^*$ and $\phi^*$ values for the currently
assumed cosmology.\\
4. The form of the LF here is that of \citet{Sau:90}, where $\alpha$ has
a slightly different definition from that in the \citet{Sch:76} LF, and the
$\sigma$ parameter effectively broadens the bright end of the LF.}
\end{deluxetable}

\begin{deluxetable}{ccccc}
\tablecaption{Measurements of $1500\,$\AA\ SFR density based on
photometric redshift estimation in the HDF and HDF-S.
 \label{lanztab}}
\tablehead{
\colhead{Redshift} & \colhead{$\log(\dot{\rho}_*)$} & \colhead{$\log(\dot{\rho}_*)$} & \colhead{$\log(\dot{\rho}_*)$} & \colhead{$\log(\dot{\rho}_*)$} \\
 & \colhead{\citet{Pas:98}} & \colhead{} & \colhead{\citet{Lan:02}} & \colhead{}
}
\startdata
 $0.25\pm0.25$ & $-1.242^{+0.310}_{-0.220}$ & $-1.807^{+0.041}_{-0.058}$ & $-1.760^{+0.030}_{-0.023}$ & $-1.713^{+0.023}_{-0.041}$ \\
 $0.75\pm0.25$ & $-1.195^{+0.230}_{-0.140}$ & $-1.631^{+0.029}_{-0.059}$ & $-1.543^{+0.017}_{-0.053}$ & $-1.526^{+0.018}_{-0.064}$ \\
 $1.25\pm0.25$ & $-1.042^{+0.240}_{-0.120}$ & $-1.526^{+0.024}_{-0.093}$ & $-1.485^{+0.029}_{-0.053}$ & $-1.456^{+0.018}_{-0.093}$ \\
 $1.75\pm0.25$ & $-1.030^{+0.240}_{-0.120}$ & $-1.508^{+0.035}_{-0.059}$ & $-1.473^{+0.023}_{-0.041}$ & $-1.391^{+0.046}_{-0.053}$ \\
 $2.25\pm0.25$ &                            & $-1.152^{+0.030}_{-0.076}$ & $-1.000^{+0.023}_{-0.064}$ & $-0.690^{+0.070}_{-0.070}$ \\
 $2.50\pm0.50$ & $-1.251^{+0.280}_{-0.210}$ & & & \\
 $2.75\pm0.25$ &                            & $-1.181^{+0.035}_{-0.082}$ & $-0.883^{+0.035}_{-0.053}$ & $-0.526^{+0.129}_{-0.175}$ \\
 $3.50\pm0.50$ & $-1.288^{+0.340}_{-0.270}$ & $-1.040^{+0.035}_{-0.094}$ & $-0.585^{+0.058}_{-0.064}$ & $-0.082^{+0.175}_{-0.316}$ \\
 $4.50\pm0.50$ & $-1.180^{+0.440}_{-0.370}$ & $-1.052^{+0.035}_{-0.094}$ & $-0.567^{+0.088}_{-0.094}$ & $-0.058^{+0.228}_{-0.415}$ \\
 $5.50\pm0.50$ & $-1.542^{+0.590}_{-0.380}$ & $-0.977^{+0.070}_{-0.093}$ & $-0.433^{+0.199}_{-0.275}$ & $0.158^{+0.357}_{-0.918}$ \\
 $8.00\pm2.00$ &                            & $-0.784^{+0.064}_{-0.111}$ & $-1.088^{+0.228}_{-0.322}$ & $0.678^{+0.386}_{-0.865}$ \\
\enddata
\end{deluxetable}

\end{document}